\documentclass[final,3p,authoryear,times]{elsarticle}

\usepackage{graphicx}

\usepackage{amssymb}

\newcommand{\jgr}{Jounral of Geophysical Research}
\newcommand{\grl}{Geophysical Research Letters}
\newcommand{\apj}{The Astrophysical Jounral}
\newcommand{\Alfven}{Alfv\'en }

\journal{Journal of Computational Physics}

\begin{document}

\begin{frontmatter}



\title{A Robust Method for Handling Low Density Regions in Hybrid Simulations
for Collisionless Plasmas}


\author{Takanobu Amano\corref{corresponding}}
\ead{amano@eps.s.u-tokyo.ac.jp}
\cortext[corresponding]{Corresponding author}

\author{Katsuaki Higashimori}
\author{Keisuke Shirakawa}
\address{Department of Earth and Planetary Science, University of Tokyo, Tokyo, 113-0033, Japan}

\begin{abstract}
A robust method to handle vacuum and near vacuum regions in hybrid simulations
for space and astrophysical plasmas is presented. The conventional hybrid
simulation model dealing with kinetic ions and a massless charge-neutralizing
electron fluid is known to be susceptible to numerical instability due to
divergence of the whistler-mode wave dispersion, as well as
division-by-density operation in regions of low density. Consequently, a pure
vacuum region is not allowed to exist in the simulation domain unless some ad
hoc technique is used. To resolve this difficulty, an alternative way to
introduce finite electron inertia effect is proposed. Contrary to the
conventional method, the proposed one introduces a correction to the electric
field rather than the magnetic field. It is shown that the generalized Ohm's
law correctly reduces to Laplace's equation in a vacuum which therefore does
not involve any numerical problems. In addition, a variable ion-to-electron
mass ratio is introduced to reduce the phase velocity of high frequency
whistler waves at low density regions so that the stability condition is
always satisfied. It is demonstrated that the proposed model is able to handle
near vacuum regions generated as a result of nonlinear self-consistent
development of the system, as well as pure vacuum regions set up at the
initial condition, without losing the advantages of the standard hybrid code.
\end{abstract}

\begin{keyword}
collisionless plasma
\sep
kinetic simulation
\sep
hybrid simulation


\end{keyword}

\end{frontmatter}


\section{Introduction}
\label{intro}

Numerical simulations have been an essential tool to investigate complicated
nonlinear phenomena occurring in space and astrophysical plasmas. Although the
conventional magnetohydrodynamics (MHD) proves itself useful to describe
macroscopic plasma dynamics even in the collisionless regime in which the mean
free path for Coulomb collisions is comparable to or larger than the system
size, it does not necessarily means that one can completely ignore important
kinetic physics. For example, it is well recognized that one must take into
account kinetic effect to understand magnetic reconnection, which has been one
of the key processes in magnetospheric physics affecting plasma transport,
driving global convection, and perhaps triggering substorms. It is now
becoming more and more popular to consider that magnetic reconnection plays a
key role in astrophysical environments as well. Another example in which
kinetic effect is central is the problem of particle acceleration in
collisionless shocks. It requires seamless treatment of both microscopic and
macroscopic physics because small-scale phenomena primarily determine the
acceleration of low energy particles (or ``injection''), while the transport
of higher energy particles is predominantly governed by characteristics of
MHD turbulence. Kinetic numerical simulations that can simultaneously deal
with both macroscopic and microscopic dynamics of the collisionless plasma are
indeed essential to investigate these important issues. Among those proposed
so far, the best numerical technique for this purpose is probably the hybrid
simulation, in which ions are treated kinetically whereas electrons are
assumed to be a massless charge-neutralizing fluid
\citep{2001sps..proc...62W,2002hmst.book.....L}.

The concept of the hybrid simulation is indeed promising in that it enables us
to access the ion dynamics, while seemingly less important but more
computationally demanding electron physics has been factored out. It has been
widely used to study elementary processes such as plasma instabilities,
magnetic reconnection, collisionless fast and slow shocks
\citep[e.g.,][]{1982JGR....87.5081L,1984JGR....89.2673W,1986JGR....91.4171T,1998JGR...103.4531N,2012JGRA..117.1220H}.
With rapidly increasing computational resources, one may now be able to use a
simulation box which is large enough to include the global scale as
well. Recently, attempts have been made to model the interaction between the
solar wind and relatively small unmagnetized and magnetized solar system
bodies by using global hybrid simulations
\citep[e.g.,][]{2002JGRA..107.1471T,2003AnGeo..21.2133K,2007GeoRL..34.5104T,2012EP&S...64..237H,2013JGRA..118.5157D}.
On the other hand, it has been well known that hybrid simulations are in
practice susceptible to numerical instability. Despite the long history of
this technique, to the authors knowledge, any fundamental solutions to this
problem has not been given. It is indeed a serious obstacle that hinders
application to many important and interesting problems in space and
astrophysical plasma physics. The primary purpose of the present paper is to
provide a practical solution to the problem of numerical stability in the
hybrid simulation. As we will see below, this can be realized by introducing a
new way to include finite electron inertia effect.

It is well known that the \Alfven wave at short wavelength comparable to ion
inertia length has dispersion due to the decoupling between ion and electron
dynamics. There thus appears the whistler mode whose frequency diverges as
$\omega \propto k^2$. This means that the maximum phase velocity in the system
increases rapidly without bound, implying numerical difficulty. This is
probably a part of the reasons for the numerical instability in hybrid
simulations. It is thus easy to expect that inclusion of finite electron
inertia can help stabilizing the simulation because the maximum phase velocity
in this case is limited by roughly the electron \Alfven speed. Even with finite
electron inertia, however, a numerical problem arises in regions of low
density. This is obviously due to the division-by-density operation needed to
calculate the electric field from ion moment quantities, which makes it
impossible to handle such (near) vacuum regions. In practice, numerical
difficulty arises even long before this limit is reached because the \Alfven
speed increases as the density decreases, imposing a severe restriction on the
simulation time step.

The method we propose in the present paper essentially resolves all these
numerical difficulties. Our strategy is also to introduce finite electron
inertia effect to limit the maximum phase velocity in the system. We argue
that the way in which the electron inertia is introduced is a key to solve the
problem. An electron inertia correction term has conventionally been
introduced to the magnetic field and its electric field counterpart is often
neglected
\citep[e.g.,][]{1998JGR...103..199K,1998JGR...103.9165S,2008JGRA..113.9204N}. By
modifying the procedure so that the correction is introduced directly to the
electric field, we show that the division-by-density operation is almost
eliminated from the simulation procedure. In addition to this, to reduce the
maximum wave phase velocity in a low density region, the ion-to-electron mass
ratio is considered to be a variable quantity. That is, the mass ratio is
reduced locally so that the CFL (Courant-Friedrichs-Lewy) condition is
automatically satisfied. We demonstrate that the proposed model implemented in
a one-dimensional (1D) hybrid simulation code can successfully follow
nonlinear evolution of the system even when extremely low density regions
appear as a result of strong instabilities. Furthermore, we also show that the
code is able to handle pure vacuum regions, as well as the interface between
vacuum and finite density plasma regions. These features suggest that the
present model is indeed very robust and will help stabilizing simulations
applied to many important problems in space and astrophysical plasmas.

The present paper is organized as follows. First, we present a simulation
model in section \ref{model}, in which a new way to introduce finite electron
inertia is discussed. Numerical implementation is explained in section
\ref{implementation}. Section \ref{Test} shows simulation results of several
test problems. Finally, summary and conclusions are given in section
\ref{summary}.

\section{Simulation Model}
\label{model}

\subsection{Standard Hybrid Model}
For the sake of completeness and to clarify the differences, we first describe
the standard hybrid model. Readers who are already familiar with the hybrid
model and its assumptions can skip this subsection. Tutorials and
comprehensive reviews of the hybrid code are found elsewhere
\citep{2001sps..proc...62W,2002hmst.book.....L}.

The basic equations used in the hybrid model are consisting of equation of
motion for individual ions and for a fluid electrons
\begin{eqnarray}
 && \frac{d \mathbf{x}_j}{dt} = \mathbf{v}_j,
 \label{eq:ion_eom1} \\
 && \frac{d \mathbf{v}_j}{dt} =
  \frac{q_j}{m_j} \left( \mathbf{E} + \frac{\mathbf{v}_j}{c} \times \mathbf{B} \right),
 \label{eq:ion_eom2} \\
 && \frac{d \mathbf{v}_e}{dt} =
  -\frac{e}{m_e} \left( \mathbf{E} + \frac{\mathbf{v}_e}{c} \times \mathbf{B} \right)
  -\frac{1}{n_e m_e} \mathbf{\nabla} \cdot \mathbf{P}_e,
  \label{eq:ele_eom}
\end{eqnarray}
where the subscript $j$ and $e$ indicate the indices for individual ions and
the electron fluid and other notations are standard.

The electromagnetic fields evolve according to the following Maxwell equations
in the Darwin approximation
\begin{eqnarray}
 && \frac{1}{c} \frac{\partial \mathbf{B}}{\partial t} =
  - \mathbf{\nabla} \times \mathbf{E},
 \label{eq:faraday} \\
 && \mathbf{\nabla} \times \mathbf{B} = \frac{4 \pi}{c} \mathbf{J},
 \label{eq:ampere}
\end{eqnarray}
and the electric charge density $\rho$ and current density $\mathbf{J}$ are
defined as
\begin{eqnarray}
 \rho &=& \sum_{s} q_s n_s  - e n_e,
  \label{eq:charge} \\
 \mathbf{J} &=& \sum_{s} q_s n_s \mathbf{V}_s
  - e n_e \mathbf{V}_e,
  \label{eq:current}
\end{eqnarray}
where $q_s, n_s, \mathbf{V}_s$ are the charge, number density and bulk
velocity of ion species $s$ calculated by taking moments of the distribution
function. Notice that there is no equation to determine the time evolution of
the electric field.

The crucial assumption in the hybrid model is the quasi-neutrality, that is,
the electrons move fast enough to cancel any charge-density fluctuations and
$\rho = 0$ is always satisfied. The electron density thus can be written by
using ion densities $n_e \approx n_i \equiv \sum_{s} q_s n_s / e$. In
addition, the electron bulk velocity may also be eliminated using Ampere's law
and the relation $\mathbf{V}_e = \mathbf{V}_i - \mathbf{J} / n_i e$ where
$\mathbf{V}_i \equiv \sum_{s} q_s n_s \mathbf{V}_s / n_e e$. Finally, since
the conventional hybrid model ignores the inertia of electron completely ($m_e
\rightarrow 0$), one can use the equation of motion for the electron fluid to
determine the electric field from given ion moment quantities and the magnetic
field. This gives the generalized Ohm's law of the form
\begin{eqnarray}
 \mathbf{E} &=&
  - \frac{\mathbf{V}_e}{c} \times \mathbf{B}
  - \frac{1}{n_i e} \mathbf{\nabla} \cdot \mathbf{P}_e, \nonumber \\
 &=&
  - \frac{\mathbf{V}_i}{c} \times \mathbf{B}
  + \frac{1}{4 \pi n_i e}
  \left( \mathbf{\nabla} \times \mathbf{B} \right) \times \mathbf{B}
  - \frac{1}{n_i e} \mathbf{\nabla} \cdot \mathbf{P}_e.
  \label{eq:ohm}
\end{eqnarray}
The second term in the right-hand side is the well-known Hall electric field
contribution. Determining the electron pressure tensor by using an appropriate
equation of state, the evolution of the system can be followed in time.

\subsection{Finite Electron Inertia}
The conventional way to include a finite electron inertia correction into the
hybrid model is to introduce the following so-called generalized
electromagnetic field $\hat\mathbf{E}$, $\hat\mathbf{B}$ defined as
\begin{eqnarray}
 \hat \mathbf{E} &=& \mathbf{E}
  - \frac{\partial}{\partial t}
  \left( \frac{c}{\omega_{pe}^2} \mathbf{\nabla} \times \mathbf{B} \right),
  \label{eq:gen_efld} \\
 \hat \mathbf{B} &=& \mathbf{B}
  + \mathbf{\nabla} \times
  \left( \frac{c^2}{\omega_{pe}^2} \mathbf{\nabla} \times \mathbf{B} \right),
  \label{eq:gen_bfld}
\end{eqnarray}
in which the terms proportional to $\mathbf{\nabla} \times \mathbf{B}$
represent electron inertia correction \citep{2002hmst.book.....L}. It is easy
to show that they exactly satisfy Faraday's law:
\begin{eqnarray}
\frac{1}{c} \frac{\partial \hat\mathbf{B}}{\partial t} =
  - \mathbf{\nabla} \times \hat\mathbf{E}.
  \label{eq:gen_faraday}
\end{eqnarray}
From the equation of motion for the electron fluid, it may be shown that
\begin{eqnarray}
 \hat\mathbf{E} = - \frac{\mathbf{V}_e}{c} \times \mathbf{B}
- \frac{1}{n_e e} \mathbf{\nabla} \cdot \mathbf{P}_e
- \frac{m_e}{e} \left( \mathbf{V}_e \cdot \mathbf{\nabla} \right) \mathbf{V}_e,
\label{eq:ohm2}
\end{eqnarray}
which is similar to the generalized Ohm's law Eq.~(\ref{eq:ohm}) but now with
the last term which also represents the correction. Note that this equation is
not exact; we have dropped the terms $\partial n_e /\partial t$, $\partial n_i
/ \partial t $, $\partial n_i \mathbf{V}_i / \partial t$, assuming ion moment
quantities do not change during the fast electron time scale.

Given the generalized electric field $\hat \mathbf{E}$, one can advance the
generalized magnetic field $\hat \mathbf{B}$ by using
Eq.~(\ref{eq:gen_faraday}). The standard electromagnetic field may then be
recovered from Eqs.~(\ref{eq:gen_efld}) and (\ref{eq:gen_bfld}). Further
simplifications are commonly adopted; for example, the electric field
correction term and electron-scale spatial variation of density are often
ignored. In this case, the magnetic field may be recovered by solving the
implicit equation
\begin{eqnarray}
 \hat \mathbf{B} &=&
  \left(1 - \frac{c^2}{\omega_{pe}^2} \mathbf{\nabla}^2 \right) \mathbf{B},
  \label{eq:gen_bfld2}
\end{eqnarray}
and $\hat \mathbf{E} \approx \mathbf{E}$ is assumed. The nice feature with
this approach is that the correction can be implemented as a post process to
the each integration step of a standard procedure.

Although the above (or similar) set of equations correctly model finite
electron inertia effect on transverse modes and have been used for a variety
of problems in space physics
\citep[e.g.,][]{1998JGR...103..199K,1998JGR...103.9165S,2008JGRA..113.9204N},
we here prefer to use a different form concerning the numerical
stability. Multiplying $n_e e$ to Eq.~(\ref{eq:ohm2}) and eliminating $\hat
\mathbf{E}$ using Eq.~(\ref{eq:gen_efld}), one obtains
\begin{eqnarray}
 \frac{1}{4 \pi} \left( \omega_{pe}^2 - c^2 \mathbf{\nabla}^2 \right) \mathbf{E}
  = \frac{e}{m_e}
  \left(
   \frac{\mathbf{J}_e}{c} \times \mathbf{B} - \mathbf{\nabla} \cdot \mathbf{P}_e
  \right)
  + \left( \mathbf{V}_e \cdot \mathbf{\nabla} \right) \mathbf{J}_e,
  \label{eq:efld_correct}
\end{eqnarray}
where $\mathbf{J}_e \equiv - e n_e \mathbf{V}_e$ is the electron current
density. In deriving this equation, $\mathbf{\nabla} \cdot \mathbf{E} \sim
O((V_A/c)^2)$ has been neglected, which is indeed a reasonable assumption.
Once the electric field is determined by solving Eq.~(\ref{eq:efld_correct}),
the magnetic field may be updated using Eq.~(\ref{eq:faraday}) without
invoking the generalized electromagnetic fields.

The present implementation obviously describes the electron scale physics
better than the conventional one because it retains the correction term for
the electric field as well. Concerning the ions dynamics, however, the effect
will be small as it affects only high frequency waves. Nevertheless, the use
of Eq.~(\ref{eq:efld_correct}) has a remarkable advantage. It is easy to
recognize that the terms in the right-hand side of Eq.~(\ref{eq:efld_correct})
are proportional to the density. (Or more precisely, they are first and second
order moments of the distribution function.) Therefore, in the limit of low
density ($n_e \approx n_i \rightarrow 0$), it correctly reduces to the
following Laplace's equation:
\begin{eqnarray}
 \mathbf{\nabla}^2 \mathbf{E} = 0,
  \label{eq:laplace}
\end{eqnarray}
implying that there is no essential difficulty with this equation in dealing
with low density (or vacuum) regions. This is reflected by the fact that the
division-by-density operation is ``almost'' eliminated in the calculation
procedure. This will be explained later in more detail.

The idea of solving Laplace's equation instead of the generalized Ohm's law
Eq.~(\ref{eq:ohm}) to obtain the electric field in low density regions is not
new. For instance, \cite{1982JCoPh..47..452H} used the same idea to allow a
vacuum region to exist in a simulation box. In this case, however, the plasma
and vacuum regions are essentially distinct and the interface between them
must somehow be determined. On the other hand, there is no need to determine
such an interface in our case. It is clear from Eq.~(\ref{eq:efld_correct})
that these two regions are naturally connected with an intermediate region in
between where the electron inertia effect dominates.

Strictly speaking, however, one must recognize the fact that dealing with such
a low density region in the hybrid model certainly violates its
assumptions. Namely, the quasi-neutrality assumption $n_e \approx n_i$ is no
longer valid in such a tenuous region because time scale associated with the
electron plasma oscillation may ultimately become comparable to the simulation
time step, and non-negligible charge density fluctuation would appear in
reality. It is thus clear that this model does not necessarily give physically
correct description of the interface between the plasma and vacuum
regions. However, with typical dynamic range of density and grid sizes in
hybrid simulations, such a region is not well resolved anyway. It is thus
rather important in practice that a code has capability to handle such regions
without numerical problems.

For later use, we rewrite Eq.~(\ref{eq:efld_correct}) into the following form
\begin{eqnarray}
 \left(
  \rho_e - \frac{m_e}{e} \frac{c^2}{4 \pi} \mathbf{\nabla}^2
  \right) \mathbf{E}
 &=& \mathbf{F}(\mathbf{B}, \rho_e, \mathbf{J}_e, \mathbf{P}_e) \nonumber \\
 &\equiv& \frac{\mathbf{J_e}}{c} \times \mathbf{B} -
  \mathbf{\nabla} \cdot \mathbf{P}_e +
  \frac{m_e}{e} (\mathbf{V}_e \cdot \mathbf{\nabla}) \mathbf{J}_e +
  \rho_e \eta \mathbf{J}
  \label{eq:efld}
\end{eqnarray}
where $\rho_e = e n_e \, (> 0)$ is the electron charge density. Here we have
introduced a finite resistivity $\eta$. It is easy to see that terms with the
$m_e/e$ factor is due to finite electron inertia which vanishes in the limit
$m_e \rightarrow 0$, and the standard generalized Ohm's law (\ref{eq:ohm}) is
recovered.

In \ref{appendix-efld}, a generalized equation for the electric field is
derived in a more systematic manner, which reduces to
Eq.~(\ref{eq:efld_correct}) in a certain limit appropriate for practical
purposes.

\subsection{Electron Pressure}
Although so far nothing has been assumed for the electron pressure tensor, in
the present study we consider only a scalar pressure $\mathbf{P}_e = P_e
\mathbf{I}$ (where $\mathbf{I}$ is a unit tensor) determined by the polytropic
equation of state for simplicity. To ensure that the pressure becomes zero in
a vacuum region $n_e \approx n_i \rightarrow 0$, we take $S \equiv P_e /
\rho_e^\gamma$ to be the independent variable where $\gamma \, (>1)$ is the
polytropic index for the electron fluid. Since it is a quantity related to the
entropy ($\propto \ln S$), its total derivative is zero in the absence of
explicit dissipation. In the presence of finite resistivity, we have
\begin{eqnarray}
 \frac{d}{dt} S =
  \frac{\partial}{\partial t} S + (\mathbf{V}_e \cdot \mathbf{\nabla}) S =
  (\gamma-1) \eta \frac{\mathbf{J}^2}{\rho_e^{\gamma}}.
  \label{eq:epressure}
\end{eqnarray}
The electron pressure $P_e$ may readily be obtained by multiplying $S$ by
$\rho_e^\gamma$. It thus vanishes in a vacuum region, consistent with
Laplace's equation Eq.~(\ref{eq:laplace}). Hereafter, the quantity $S$ is
called the electron entropy although it is not so in a strict sense.

Note that the application of our equation for the electric field is not
restricted to the specific model of the electron pressure tensor. Extension to
any tensor electron pressure models proposed previously, such as those used to
study collisionless magnetic reconnection
\citep[e.g.,][]{1994JGR....9911177H,1998JGR...103..199K}, is straightforward.

\section{Numerical Implementation}
\label{implementation}

In this section, implementation of the proposed model to a 1D code is
described. We think that the scheme given here is just an example and
different methods may also be used and extension to multidimensions should be
straightforward because the essential difference from the standard hybrid code
is only the way in which the electric field is determined.

\subsection{Time Integration}
The standard Buneman-Boris integration is used to calculate particle
trajectories. The particle positions and velocities are defined at the integer
and half-integer time steps ($\mathbf{x}_j^{n}$, $\mathbf{v}_j^{n+1/2}$),
respectively. Accordingly, the electromagnetic field is defined at the integer
time step $\mathbf{E}^n$, $\mathbf{B}^n$.

We use the following iterative algorithm of \cite{1989JCoPh..84..279H} for
time integration of the induction equation.
\begin{eqnarray}
 \mathbf{B}^{n+1/2} &=& \mathbf{B}^n - \frac{c \Delta t}{2} \mathbf{\nabla}
  \times \mathbf{E}^{n+1/2} \\
 \left( \rho_{e}^{n+1/2} - \frac{m_e c^2}{4 \pi e} \mathbf{\nabla}^2 \right)
  E_{k}^{n+1/2} &=&
  F_{k}(\mathbf{B}^{n+1/2}, \rho_e^{n+1/2},
  \mathbf{J}_e^{n+1/2}, \mathbf{P}_e^{n+1/2}), \\
 \rho_{e}^{n+1/2} E_x^{n+1/2} &=&
  F_x(\mathbf{B}^{n+1/2}, \rho_e^{n+1/2}, \mathbf{J}_e^{n+1/2},
  \mathbf{P}_e^{n+1/2}),
\end{eqnarray}
where $k = y, z$. Notice that the longitudinal component ($E_x$) does not have
the Laplacian correction term. This comes from the fact that the longitudinal
and transverse components decouple in 1D, and the correction must operate only
to the transverse component. In multidimensional simulations, one may simply
introduce the correction to all the components because (1) $\mathbf{\nabla}
\cdot \mathbf{E}$ is small and (2) it is not easy (unless one solves Poisson's
equation) to decompose the field into the transverse and longitudinal
components.

We define the electron entropy (or equivalently pressure) at the half time step
$S^{n+1/2}$, which is advanced by using the electron velocity defined at the
full time step $\mathbf{V}_e^{n}$ as follows
\begin{eqnarray}
 S^{n+1/2} = S^{n-1/2} - \Delta t
  \left[
  \left(
   \mathbf{V}_e^{n} \cdot \mathbf{\nabla}
  \right) S^{n-1/2} +
  (\gamma-1) \eta \frac{(\mathbf{J}^{n})^2}{(\rho_e^{n})^{\gamma}}
  \right],
  \label{eq:electron-entropy}
\end{eqnarray}
and then used to determine the electric field $\mathbf{E}^{n+1/2}$. The
iteration is typically performed until relative error of the electric field
becomes smaller than $10^{-3}$ at all grid points. Although this formally
looks an implicit scheme, we find it is not stable when the CFL condition
defined for whistler wave phase velocity is violated.

The electromagnetic field at the next time step is then determined as follows
\begin{eqnarray}
 \mathbf{B}^{n+1} &=& - \mathbf{B}^{n} + 2 \mathbf{B}^{n+1/2} \\
 \mathbf{E}^{n+1} &=& - \frac{1}{2} \mathbf{E}^{n-1/2} + \frac{3}{2} \mathbf{E}^{n+1/2}.
  \label{eq:efd_next}
\end{eqnarray}
Notice that the electric field $\mathbf{E}^{n+1}$ is estimated from those
defined at half time steps. We find that naive use of the relation
$\mathbf{E}^{n+1/2} = (\mathbf{E}^{n+1} + \mathbf{E}^{n})/2$ results in
producing high frequency aliasing noise in the electric field spectrum, which
completely vanishes when Eq.~(\ref{eq:efd_next}) is used instead. This may be
understood by the fact that the value of $\mathbf{E}^{n}$ is not well
constrained because adding an arbitrary amount to $\mathbf{E}^{n}$ and
subtracting the same amount from $\mathbf{E}^{n+1}$ does not change
$\mathbf{E}^{n+1/2}$. Since $\mathbf{E}^{n+1/2}$ is well determined by the
above iteration procedure, Eq.~(\ref{eq:efd_next}) better estimates the
electric field at the next step.

\subsection{Spatial Discretization and Electron Inertia Correction}
We use the standard staggered mesh for the electromagnetic field
$\mathbf{B}_{i+1/2}$, $\mathbf{E}_{i}$ with a constant grid spacing $\Delta
x$. To be consistent with this, ion moment quantities and the electron entropy
are defined at the integer grid points: $\rho_{e,i}$, $\mathbf{J}_{e,i}$,
$S_{i}$. The second-order central finite difference is used for approximation
of spatial derivatives except for the electron entropy equation
(\ref{eq:electron-entropy}) which is solved by the first-order upwind
scheme. It is well known that the staggered mesh can be extended to
multidimensions and it guarantees $\mathbf{\nabla} \cdot \mathbf{B} = 0$
within machine epsilon.

We solve the implicit equation (\ref{eq:efld}) for the electric field in an
iterative manner. The right-hand-side $\mathbf{F}$ calculated from the moment
quantities and magnetic field at each grid point is used as a source term for
solving the equation. The second-order finite difference approximation to the
Laplacian operator reduces it to a tridiagonal matrix equation in 1D
\begin{eqnarray}
 \left(
  -\epsilon E_{k,i-1} +
  \left(\rho_{e,i} + 2 \epsilon \right) E_{k,i} +
  -\epsilon E_{k,i+1}
 \right) = F_{k,i},
\end{eqnarray}
where $k = y, z$ and $\epsilon = m_e c^2 / 4 \pi e \Delta x^2$. Notice that
$\epsilon/\rho_e = (c/\omega_{pe} / \Delta x)^2 \lesssim 1$ in practice,
meaning that the matrix is diagonally dominant and is relatively easy to
invert. In this study, we use the simple symmetric Gauss-Seidel method to
solve the matrix equation which is very easy to implement. Although its
convergence is known to be slow, experience has shown that only a few
iterations are typically sufficient.  In general, with higher order
discretization and/or in multidimensions, it becomes a band matrix. The
diagonally dominant property, however, does not change because it is
determined by the fact that the electron inertia is merely a small
correction. The situation obviously changes when the grid size is chosen to be
small enough to resolve the electron inertia length $c/\omega_{pe}$ to take
into account the electron-scale physics more rigorously. In addition, in the
case where a pure vacuum region exists in the simulation domain as is treated
in one of the test problems discussed below, the implicit equation essentially
reduces to Laplace's equation. In such a case, it is better to use a more
sophisticated iterative matrix solver for faster convergence.

It is worth noting that, once the source term is given, the
division-by-density operation is not anymore needed to invert the matrix,
because the diagonal coefficients (whose inverses are needed) of the matrix is
$\rho_{e,i} + 2 \epsilon$, rather than $\rho_{e,i}$. However, we must mention
that the calculation of $\mathbf{V}_e = \mathbf{J}_e / \rho_e$ cannot
completely be avoided from the numerical procedure: It appears at the third
term of the source term of Eq.~(\ref{eq:efld}) and the equation for the
electron entropy Eq.~(\ref{eq:epressure}). Nevertheless, these terms do not
pose a serious numerical problem in practice because they represent the
Doppler shift of the waves into the electron-fluid rest frame whenever it is
well-defined, while on the other hand, in a vacuum region obviously it cannot
be defined. For the present purpose, we redefine the electron velocity as
\begin{eqnarray}
 \mathbf{V}_e = \frac{\mathbf{J}_e}{\max(\rho_e, \rho_{e,min})},
\end{eqnarray}
where the minimum density $\rho_{e,min}$ is chosen to correspond to the
one-count level in the present study. This makes sure that the electron
velocity does not diverge in a vacuum region whereas the modification does not
affect results in a cell containing more than one particle. We find this
simple fix, combined with the variable mass ratio technique as explained
below, is sufficient to keep track of the simulation without numerical
problems even for highly nonlinear problems in which near vacuum regions
appear in an unexpected manner.

\subsection{Variable Mass Ratio}
The numerical schemes described so far have been shown to be successful when
the time step is sufficiently small. However, it may become unstable in highly
nonlinear problems in which regions of extremely low density appear as a
result of self-consistent time evolution and consequently the maximum phase
velocity of the system violates the CFL condition. A much smaller time step is
therefore needed for stability, but it is often impractical. We simply try to
stabilize the simulation by numerical means at the expense of correct physics
whenever they appear in the simulation box.

For a cold plasma, the maximum wave phase velocity in the system is determined
from the whistler mode dispersion relation as
\begin{eqnarray}
 v_{p, max} \simeq \frac{1}{2} \frac{B}{\sqrt{4 \pi n_e m_e}} =
  \frac{1}{2} V_{A,e}
\end{eqnarray}
where $V_{A,e}$ is the electron \Alfven velocity. This upper bound on the phase
velocity does not exist in the limit $m_e \rightarrow 0$ because the electron
\Alfven velocity is proportional to $\sqrt{m_i/m_e}$.

In general, numerical stability for an explicit scheme at least requires the
Courant number defined with respect to the maximum phase velocity to be less
than unity: $v_{p, max} \Delta t / \Delta x \leq 1$. The phase velocity must
therefore be numerically reduced for stability in low density regions. Our
strategy here is to consider the mass ratio as a variable quantity. By locally
and temporarily modifying the mass ratio, the maximum phase velocity can be
reduced so that the CFL condition is always satisfied. More specifically, one
may use a modified electron mass $m_e'$ defined as
\begin{eqnarray}
 \frac{m_e'}{m_i} = \max
  \left(
   \frac{m_e}{m_i},
   V_{A}^2
   \left(\frac{\Delta t}{2 \alpha \Delta x}\right)^2
  \right)
  \label{eq:modified_mass}
\end{eqnarray}
instead of the physical electron mass $m_e$. Here $V_A = B/\sqrt{4 \pi n_i
m_i}$ is the \Alfven speed calculated by using the local density and magnetic
field, and $\alpha$ is the maximum allowed Courant number. In the present
paper, we always choose $\alpha = 1/2$ for safety.

It is important to mention that although this modifies the dispersion relation
of whistler waves, low frequency \Alfven waves are not affected as far as
$m_e/m_i \ll 1$ is satisfied. In contrast, if one imposes a floor value in
density to reduce the phase velocity, it is modified even in the MHD limit. In
addition, since the phase velocity diverges at short wavelength in the limit
$m_e \rightarrow 0$, the floor value must be chosen much larger than the case
with finite electron inertia. It must be pointed out that the electron inertia
length $c/\omega_{pe}$ introduced in hybrid and Hall-MHD models is usually
treated as if it were a constant even when the density may change
substantially
\citep{1998JGR...103..199K,1998JGR...103.9165S,2008JGRA..113.9204N}. This
corresponds to changing the mass ratio to compensate the density
variations. Our treatment is similar, but much better than this because it may
change only in the limited region and time in which the stability condition is
no longer satisfied. In any case, modification of the finite electron mass
will not influence the simulation results as far as the electron inertia
scale is not appropriately resolved.

We emphasize that when one is primarily interested in the ion dynamics, a
finite electron mass may be seen as an artificial parameter for numerical
stability rather than physical, which does not affect the simulation results
by assumption. This will hold in most of situations where the hybrid
simulation applies and the electron inertia effect is not expected to be
important. Otherwise, one must employ a more fundamental model taking into
account rigorous electron-scale physics. In \ref{appendix-me}, validity range
of such variable mass ratio technique is discussed.

\section{Test Problems}
\label{Test}

In this section, we discuss simulation results for several test problems that
demonstrate the robustness of our new method. In the present paper, we fix the
number of iteration for the electron inertia correction (symmetric
Gauss-Seidel iteration) to 2, while the iteration to determine the electric
field (i.e., the Horowitz iteration loop) continues until the relative error
of electric fields becomes less than $10^{-3}$. Note that we have confirmed
that the number of iteration for the electron inertia correction does not
change the result significantly.

In the following, unless otherwise stated, time and space are respectively
normalized to the inverse ion cyclotron frequency $1/\Omega_{ci}$ and the ion
inertia length $V_{A}/\Omega_{ci}$ defined for the average density and
magnetic field. The speed of light is held fixed to $V_A/c = 10^{-4}$ in all
the simulation runs presented below. The resistivity is assumed to be zero
except for examples shown in section 4.4. The number of grids, number of
particle per cell are denoted by $N_x$, $N_{ppc}$, respectively. The periodic
boundary condition is always used, and the velocity distribution is
initialized by isotropic Maxwellian with a given temperature.

\subsection{Linear Dispersion Relation}
\begin{figure}[t]
 \includegraphics[width=14.0cm]{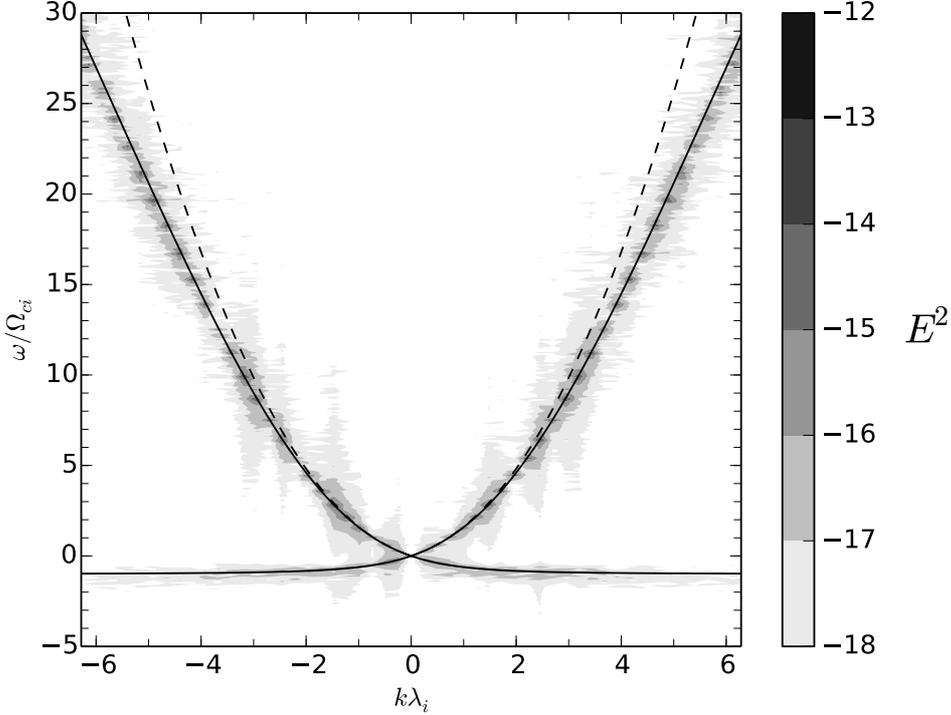}

 \caption{Power spectral density (gray scale) obtained from the simulation for
 a homogeneous plasma. No explicit perturbations are added at the initial
 condition. The solid line represents the dispersion relations of circularly
 polarized electromagnetic waves calculated for a cold plasma. The dashed line
 shows the dispersion relation corresponding to the Hall-MHD ($m_e \rightarrow
 0$).}

 \label{fig:linear-dispersion}
\end{figure}

We have tested whether our newly developed code can reproduce theoretical
linear dispersion relation for a homogeneous plasma with finite electron
inertia effect. The simulations are performed without any explicit
perturbations and the system evolves solely from thermal noise. The code is
verified with various grid spacings and time steps, number of particles per
cell, as well as physical parameters such as mass ratio, plasma beta, etc. One
of the examples is shown in Fig.~\ref{fig:linear-dispersion}, which displays
the $\omega {\rm - } k$ diagram of the transverse electric field obtained with
a mass ratio $m_i/m_e = 100$, temperature ratio $T_i/T_e = 1$, and plasma beta
$\beta_i = 10^{-2}$. Here, the constant background magnetic field $B_0$ is
imposed along the $x$ direction and the electric field is normalized to $V_A
B_0 / c$ accordingly. Other simulation parameters are as follows: $\Delta t =
5 \times 10^{-3}$, $\Delta x = 0.1$, $N_x = 256$, $N_{ppc} = 64$. In this
plot, the positive (negative) frequency represents right-hand (left-hand)
circularly polarized mode, and similarly signs of the wavenumber indicate
different helicities. We see that the agreement between the simulation result
and theoretical dispersion relation shown with the solid lines is very
good. Notice that, in this figure, the effect of electron inertia appears at
$k \lambda_i \gtrsim 3$, beyond which the phase velocity is reduced relative
to the Hall-MHD dispersion relation shown in the dashed line. We have
confirmed that the total energy is in general very well conserved. In this
particular run, the error is less than $\sim 2 \times 10^{-3}$ \%.

\subsection{Electromagnetic Ion Beam Instability}
In this section, we discuss simulation results for the resonant
electromagnetic ion beam instability which is one of the standard test
problems for a conventional hybrid code. The purpose of this test is to show
that the present method does not introduce any additional numerical
difficulties when applied to problems that can be treated by the standard
hybrid model.

\begin{figure}[t]
 \includegraphics[width=14.0cm]{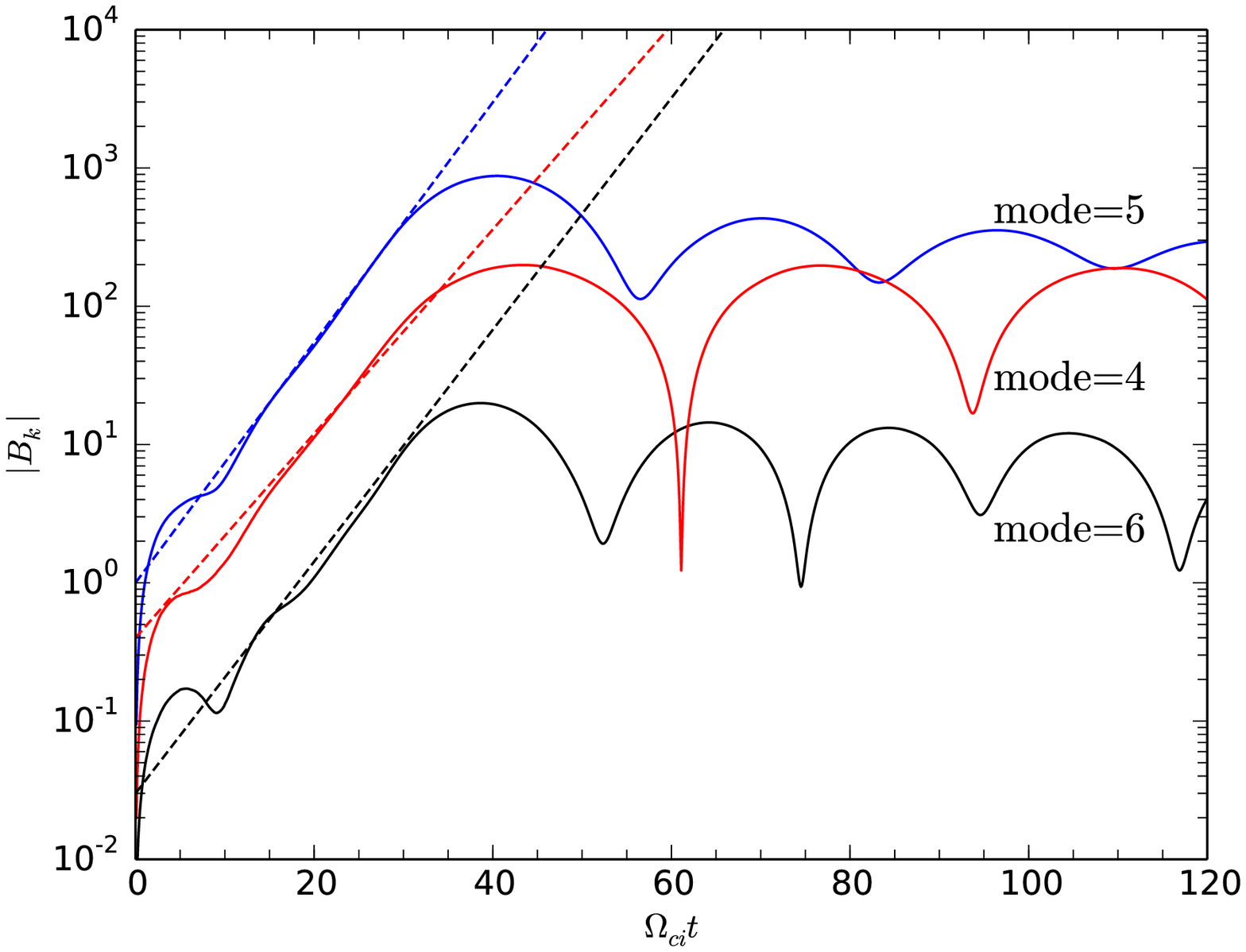}

 \caption{Time evolution of mode amplitudes expected to grow due to the
 electromagnetic ion beam instability. The different colors represent
 different modes (mode numbers $4, 5, 6$). The dashed lines indicate
 theoretical linear growth rates for corresponding modes shown in the solid
 lines.}

 \label{fig:beam-linear-growth}
\end{figure}

\begin{figure}[t]
 \includegraphics[width=14.0cm]{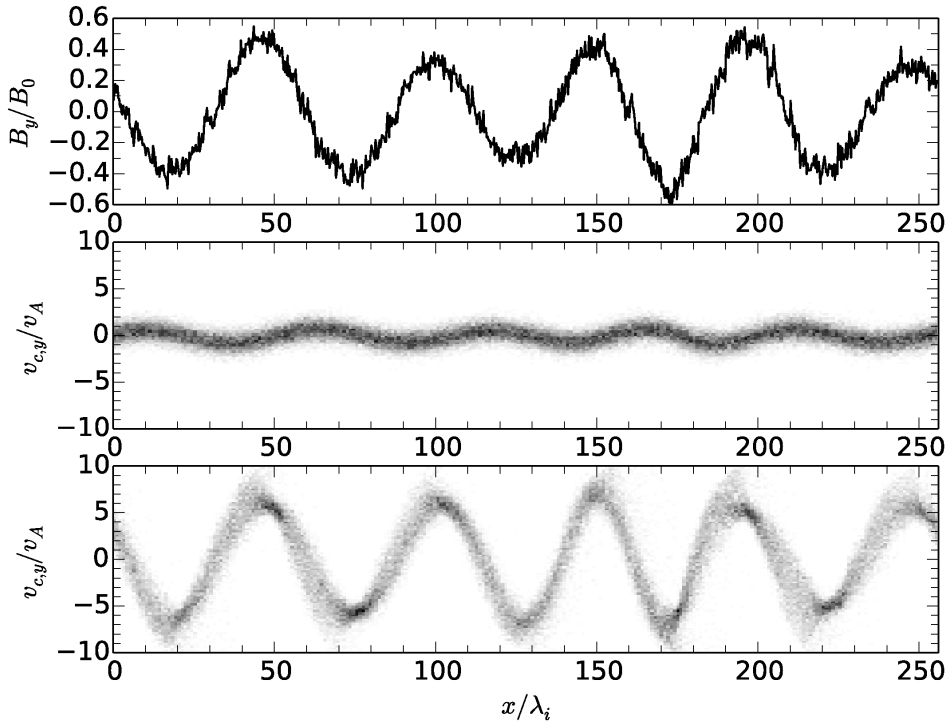}

 \caption{Snapshot of magnetic field $B_y$ (top) and transverse ion phase
 space diagram for core (middle) and beam (bottom) components just before the
 saturation.}

 \label{fig:beam-snapshot}
\end{figure}

\begin{figure}[t]
 \includegraphics[width=14.0cm]{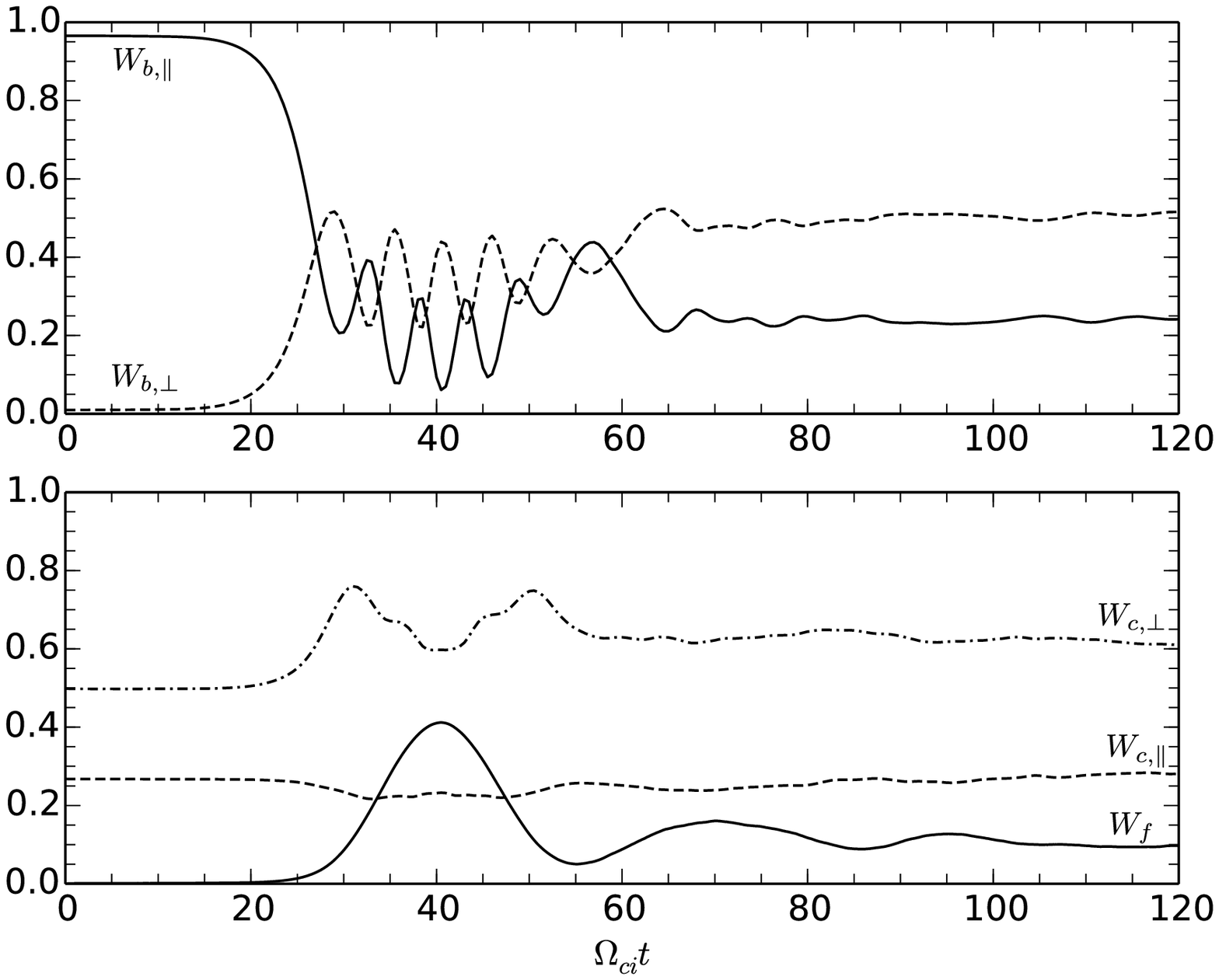}

 \caption{Time history for parallel and perpendicular particle energies for
 the beam $W_{b,\parallel}$, $W_{b,\perp} $(top), core ions $W_{c,\parallel}$,
 $W_{c,\perp}$, and the magnetic field energy $W_f$ (bottom).}

 \label{fig:beam-history}
\end{figure}

The simulation setup is very similar to \cite{1984JGR....89.2673W}. We choose
the relative beam density $n_b/n_0 = 0.02$, and bulk velocity $V_b/V_A = 10$
streaming parallel to the ambient magnetic field, the plasma beta for core
ions $\beta_c = 1.0$, beam ions $\beta_b = 1.0$, and electrons $\beta_e =
0.1$. The ion to electron mass ratio is chosen to be $m_i/m_e = 100$. In this
case, it is easy to find from linear analysis that right-hand circularly
polarized electromagnetic waves propagating parallel to the magnetic field are
unstable due to cyclotron resonant interaction with the beam ions. The initial
magnetic field is thus taken to be along the $x$
direction. Fig.~\ref{fig:beam-linear-growth} displays the time evolution of
the mode amplitudes that are expected to grow due to this instability. In this
simulation, we use $\Delta t = 0.01$, $\Delta x = 0.25$, $N_x = 1024$,
$N_{ppc} = 64$. It is clear that the simulation results agree very well with
the linear theory whose growth rates are shown in dashed
lines. Fig.~\ref{fig:beam-snapshot} shows a snapshot of the magnetic field,
and the ion phase space diagram for both beam and core components just before
the saturation. The transverse velocities of beam ions are strongly affected
by the excited large-amplitude wave whereas the core ions are modulated only
slightly. This feature is consistent with the fact that the instability is
excited by the resonance between the wave and beam ions.
Fig.~\ref{fig:beam-history} shows time evolution of parallel and perpendicular
energies for both beam and core ions, as well as the magnetic field
energy. One can see that the beam parallel energy is substantially reduced as
a result of the instability and transferred to the perpendicular energy of the
beam and core components through pitch-angle scattering.

The above linear and nonlinear development of the instability is consistent
with previous studies, and thus confirms that the new model can reproduce the
standard hybrid simulation results. Note that, in this simulation, the grid
size is always much larger than the electron inertia length during the whole
run and the electron inertia effect is an unimportant small
correction. Indeed, the result without the correction appears almost exactly
the same. It is important to mention that the inclusion of the electron
inertia effect does not impose any numerical difficulties in application of
our method to problems where the electron inertia does not play a role.

\subsection{Decay Instability}
Now we consider an example in which near vacuum (or extremely low density)
regions appear as a result of nonlinear and self-consistent development of the
system. The standard hybrid method will not be able to keep track of such a
simulation owing to its limit on handling low density regions. In contrast, we
demonstrate that the present model is free from such difficulty.

To show this, we choose a parametric instability of a large-amplitude
circularly polarized \Alfven wave
\citep{1978ApJ...219..700G,1986JGR....91.5617W,1986JGR....91.4171T}. Although
the \Alfven wave is an exact solution of MHD equations even for finite
amplitude, it is known to be unstable against perturbations and will decay
through the excitation of other waves (daughter waves). Specifically, the
decay instability is a process occurring in a low beta plasma that excites a
forward-propagating ion-acoustic wave and a backward-propagating \Alfven
wave. The simulation parameters are chosen as follows. The parent
large-amplitude wave is on the R-mode \Alfven/whistler branch propagating along
the ambient magnetic field (taken along the $x$ direction) with frequency and
wavenumber of $(\omega_0, k_0) = (0.215, 0.196)$ and the amplitude is $B_p/B_0
= 0.5$ (i.e., 50\% of the background magnetic field). The plasma beta (for
both ions and electrons) and ion-to-electron mass ratio are $\beta_i = \beta_e
= 10^{-2}$ and $m_i/m_e = 100$, respectively. Other parameters are $\Delta t =
0.01$, $\Delta x = 0.5$, $N_x = 512$, $N_{ppc} = 64$.

\begin{figure}[t]
 \includegraphics[width=14.0cm]{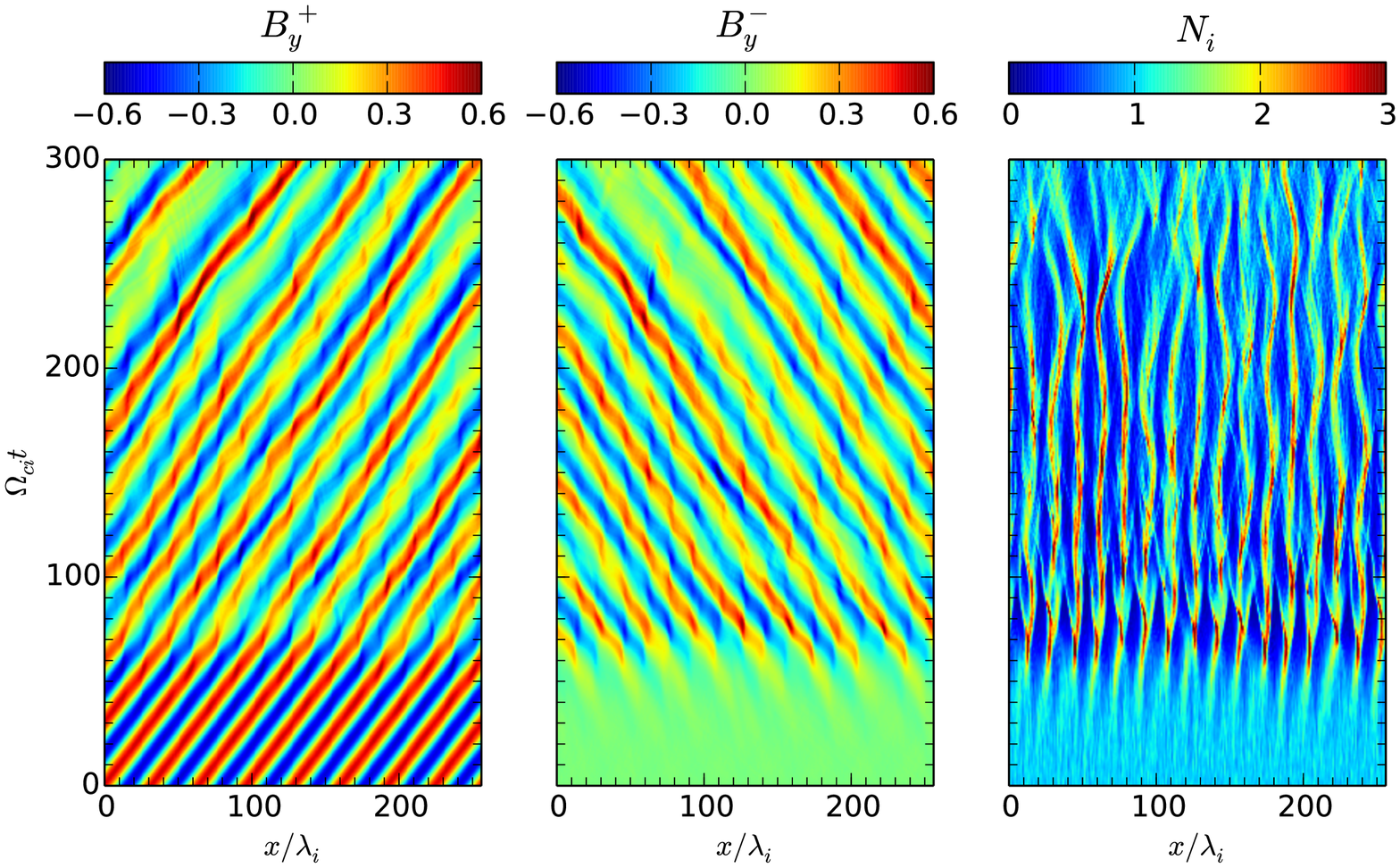}

 \caption{Time evolution of decay instability of \Alfven wave. The left and
 center panels show $B_y^+$ and $B_y^-$ which are calculated using Fourier
 decomposition of the raw $B_y$ into different helicities. The right panel
 shows density fluctuations normalized to the initial uniform density.}

 \label{fig:decay-evolution}
\end{figure}

Time evolution of the decay instability is summarized in
Fig.~\ref{fig:decay-evolution}. In this figure, the transverse magnetic field
$B_y$ is decomposed into different helicities (denoted as $B_y^+$ and
$B_y^-$), and are shown in the left and center panels, respectively. Note that
the $B_y^+$ ($B_y^-$) component includes R-mode (L-mode) waves propagating to
the right and L-mode (R-mode) wave propagating to the left. (See
\cite{1986JGR....91.4171T} for technical details.) The density fluctuations
are also shown in the right panel. The parent wave is a R-mode wave
propagating in the positive $x$ direction with a mode number of 8, which can
be easily identified in the left panel at the initial stage. The growth of the
instability is seen as the development of large-amplitude density
fluctuations. A backward-propagating \Alfven wave is then excited associated
with this. For this particular run, the amplitude of density fluctuations is
found to be substantial ($N_{i} / N_{i,0} \gtrsim 1$) at the saturation stage.

\begin{figure}[t]
 \includegraphics[width=14.0cm]{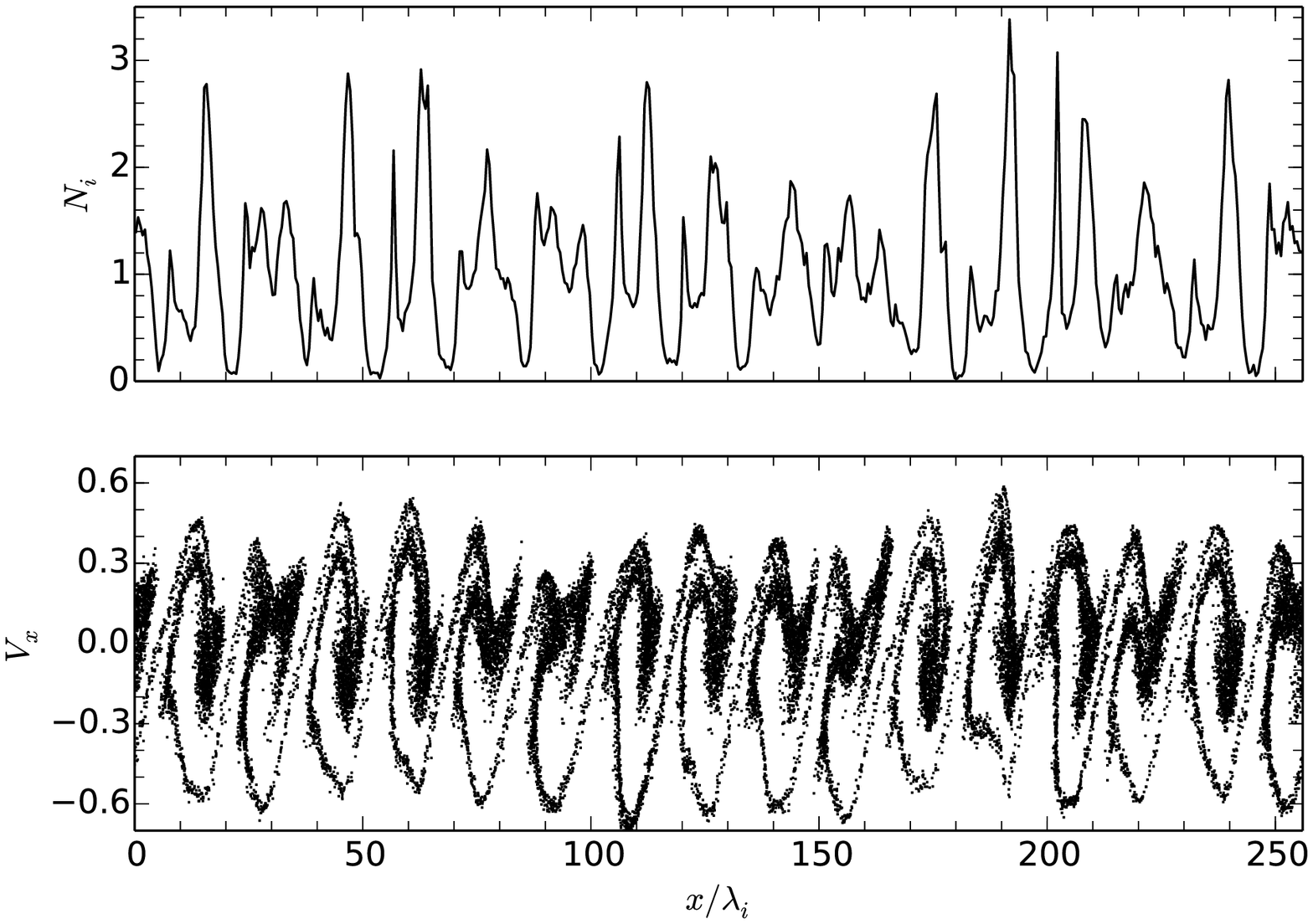}

 \caption{Snapshot of density (top) and ion phase space diagram (bottom) for
 decay instability of \Alfven wave around the saturation $\Omega_{ci} t =
 92.0$.}

 \label{fig:decay-snapshot}
\end{figure}

Fig.~\ref{fig:decay-snapshot} displays the snapshot of density and ion phase
space diagram around the saturation $\Omega_{ci} t = 92.0$. One can see that
clear ion phase-space holes are formed due to the trapping of ions by an
electrostatic potential produced by large-amplitude ion-acoustic waves as was
found by earlier studies \citep{1986JGR....91.4171T}. The large-amplitude
density fluctuations are associated with the trapped ion dynamics. The minimum
density (during the whole run) goes down to $\sim10^{-2}$ relative to the
initial density, which is comparable to the one-count level ($1/N_{ppc} \sim
0.016$) for the simulation parameters. Thanks to the new method to determine
the electric field as well as the variable mass ratio technique, the
appearance of such extremely low density regions does not lead to collapse of
the simulation. Note that we have confirmed that a fixed mass ratio makes
simulations numerically unstable even if the same equation is used to
determine the electric field. Furthermore, even densities below the one-count
level do not lead to any numerical problems. Indeed, a simulation with a lower
beta ($\beta = 10^{-3}$) with all other parameters fixed is also successful,
in which the minimum density becomes as low as $\sim 10^{-3}$, i.e., well
below the one-count level. This demonstrates the robustness of our model in
handling low density (or near vacuum) regions which may appear in an
unexpected manner due to nonlinear development of the system. This is a clear
advantage over the standard hybrid code.

\subsection{Plasma Expansion to Vacuum}
Finally, we demonstrate that the present model is able to handle a pure vacuum
as well as the interface between plasma and vacuum regions in a seamless
manner. One of the examples of such situations occurring in space is the
interaction between the solar wind and the Moon
\citep[e.g.,][]{2012EP&S...64..237H}. Since the solar wind plasma is obscured,
the plasma density is substantially depleted behind the obstacle and a wake
region appears which is essentially a vacuum region. The solar wind plasma
gradually intrude into the wake owing to a finite thermal velocity and the
region will be filled with the plasma again far downstream of the
obstacle. Assuming steady state, this refilling process of the wake may be
approximately modeled by 1D expansion of a plasma into a pure vacuum region
\citep{1998JGR...10323653F,2001PhPl....8.4551B}, which is simulated here.

\begin{figure}[t]
 \includegraphics[width=14.0cm]{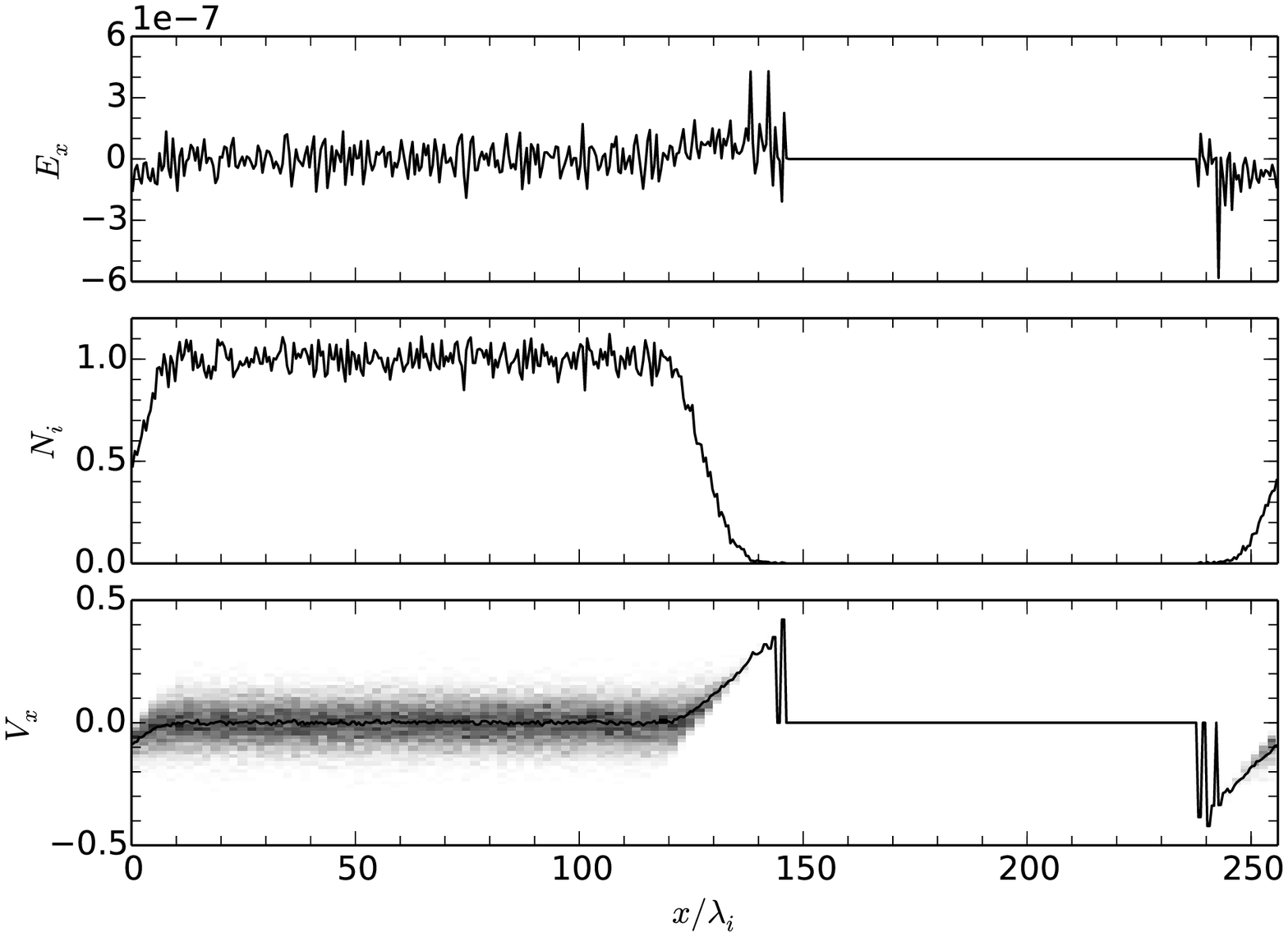}

 \caption{Snapshot of plasma expansion to vacuum at $\Omega_{ci} t =
 48.0$. The longitudinal electric field (top), density (middle), ion phase
 space diagram (gray scale) and $x$ component of the bulk velocity $V_x$
 (bottom) are shown. In a pure vacuum region where ion density is exactly
 zero, the bulk velocity is also set to zero.}

 \label{fig:expansion-homogeneous}
\end{figure}

Initially, the system is divided into two regions: the left and right regions
respectively correspond to the plasma and vacuum regions. The plasma then
freely expands into the vacuum region with their thermal velocity. This can be
clearly seen in Fig.~\ref{fig:expansion-homogeneous} showing a snapshot of a
typical simulation. The plasma region is initially uniform and is
characterized by $\beta_i = \beta_e = 10^{-2}$, and $m_i/m_e = 100$. The
uniform magnetic field $B_0$ parallel to the $x$ direction is imposed. Other
parameters $\Delta t = 0.01$, $\Delta x = 0.5$, $N_x = 512$ and $N_{ppc} =
128$ for the plasma region are used. Note that, in this section, the inertia
length and \Alfven velocity (i.e., normalizations) are defined with the
average density over the entire simulation box.

Since the magnetic field is along $x$ direction, particles in the left-hand
side of the box with positive velocity can freely propagate to the vacuum
region. This free streaming signature can be seen in the ion phase space
diagram. In addition, the longitudinal electric field $E_x$ in the interface
region is slightly positive due to the pressure gradient ($x \sim 130{\rm
-}140$), and the ions at the leading edge are accelerated toward the vacuum, a
feature consistent with previous particle-in-cell simulations
\citep{1998JGR...10323653F,2001PhPl....8.4551B}. We see that the code can keep
track of the evolution without any numerical instabilities even in the
presence of a pure vacuum region. No ad hoc technique is needed in handling
the interface region.

There is concern about handling vacuum regions with hybrid codes because a
finite current density may numerically arise even in the absence of current
carriers. This is because the total current is calculated from the magnetic
field and is nothing to do with the plasma density. This clearly contradicts
with the basic assumption of the hybrid model. To let the system quickly
relaxes to a state consistent with the assumption, previous studies have
introduced a large resistivity in low density regions
\citep{1980JCoPh..38..378H,2013ASPC..474..202H}. The resulting equation for
the magnetic field is a diffusion equation with its coefficient proportional
to the resistivity, and the steady state solution is given by a potential
magnetic field $\mathbf{\nabla}^2 \mathbf{B} = 0$. The problem with this
approach is that the large resistivity imposes a severe restriction on the
time step for an explicit time integration scheme. In contrast to this, our
method can better handle this issue.

In the region of our interest, the plasma density $\rho_e$ approaches to zero,
whereas we need a large resistivity $\eta$. We may thus assume that $\eta
\rho_e$ remains finite. Then for a sufficiently low density region, the equation
for the electric field is reduced to
\begin{eqnarray}
 \mathbf{\nabla}^2 \mathbf{E} = - \frac{\omega_{pe}^2}{c^2} \eta \mathbf{J}.
\end{eqnarray}
In this case, by taking rotation of the induction equation, one sees
\begin{eqnarray}
 \frac{\partial}{\partial t}
  \left( \mathbf{\nabla} \times \mathbf{B} \right)
  = - \eta \frac{\omega_{pe}^2}{4 \pi}
  \left( \mathbf{\nabla} \times \mathbf{B} \right)
\end{eqnarray}
which means that the total current in such a region decays exponentially with
a damping rate of $\eta \omega_{pe}^2/4\pi$. This equation is clearly a pure
damping equation requiring only local information. Therefore, the numerical
stability criterion is very much relaxed as compared to the diffusion equation
which involves spatial derivatives. Practically, one can use a large
resistivity such that the decay time becomes on the order of the simulation
time step. Since the total current vanishes as a result, the final state will
be given by a potential magnetic field that is equivalent to the diffusion
equation approach. Namely, even if a non-zero current density develops for
whatever reason, one can enforce it to decay very rapidly within a few time
steps by appropriately choosing the resistivity. We have implemented this
damping by setting the resistivity as follows:
\begin{eqnarray}
 \eta \rho_e = \frac{1}{2}
  \left(
   \gamma_{max} \frac{m_e}{e} - \eta_0 \rho_e
  \right)
  \left(
   1 - \tanh \left( \frac{\rho_e - \rho_{min}}{\sigma} \right)
  \right) + \eta_0 \rho_e,
\end{eqnarray}
where $\eta_0$ is the background uniform resistivity (chosen to be zero in
this study), $\gamma_{max}$ is the damping rate in vacuum, $\rho_{min}$ is
density corresponding to one particle per cell and we choose $\sigma =
\rho_{min}/2$. This form of resistivity smoothly connects from the uniform
background in the plasma region to the vacuum where the damping rate is chosen
to be $\gamma_{max} = 1/\Delta t$, and ensures that the large resistive
damping operates only in near and pure vacuum regions where $\rho_e \lesssim
\rho_{min}$.

\begin{figure}[t]
 \includegraphics[width=14.0cm]{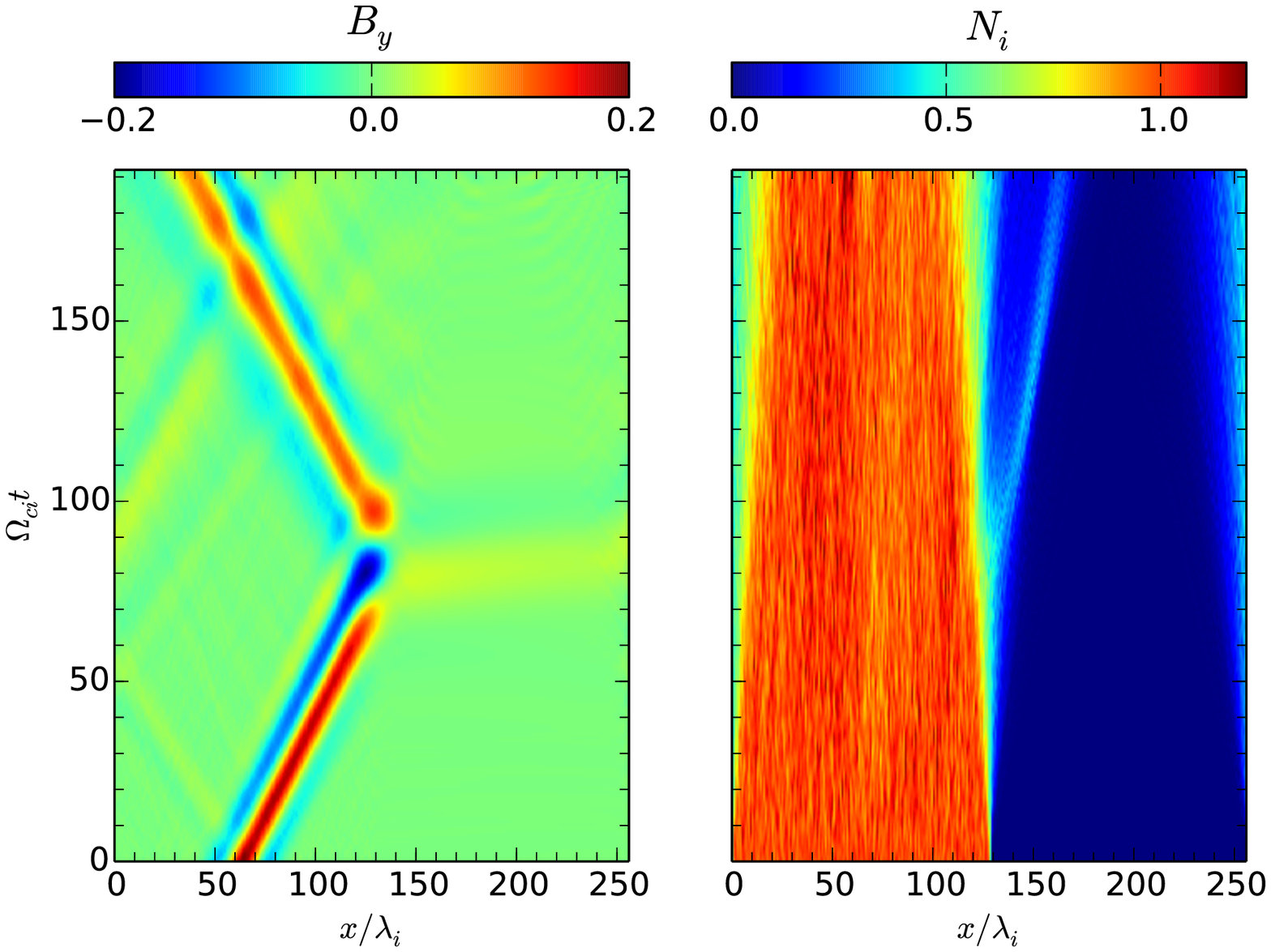}

 \caption{Time evolution of plasma expansion to vacuum with localized wave
 packet. The transverse magnetic field $B_y$ (left) and density (right) are
 shown.}

 \label{fig:expansion-wave-evolution}
\end{figure}

\begin{figure}[t]
 \includegraphics[width=14.0cm]{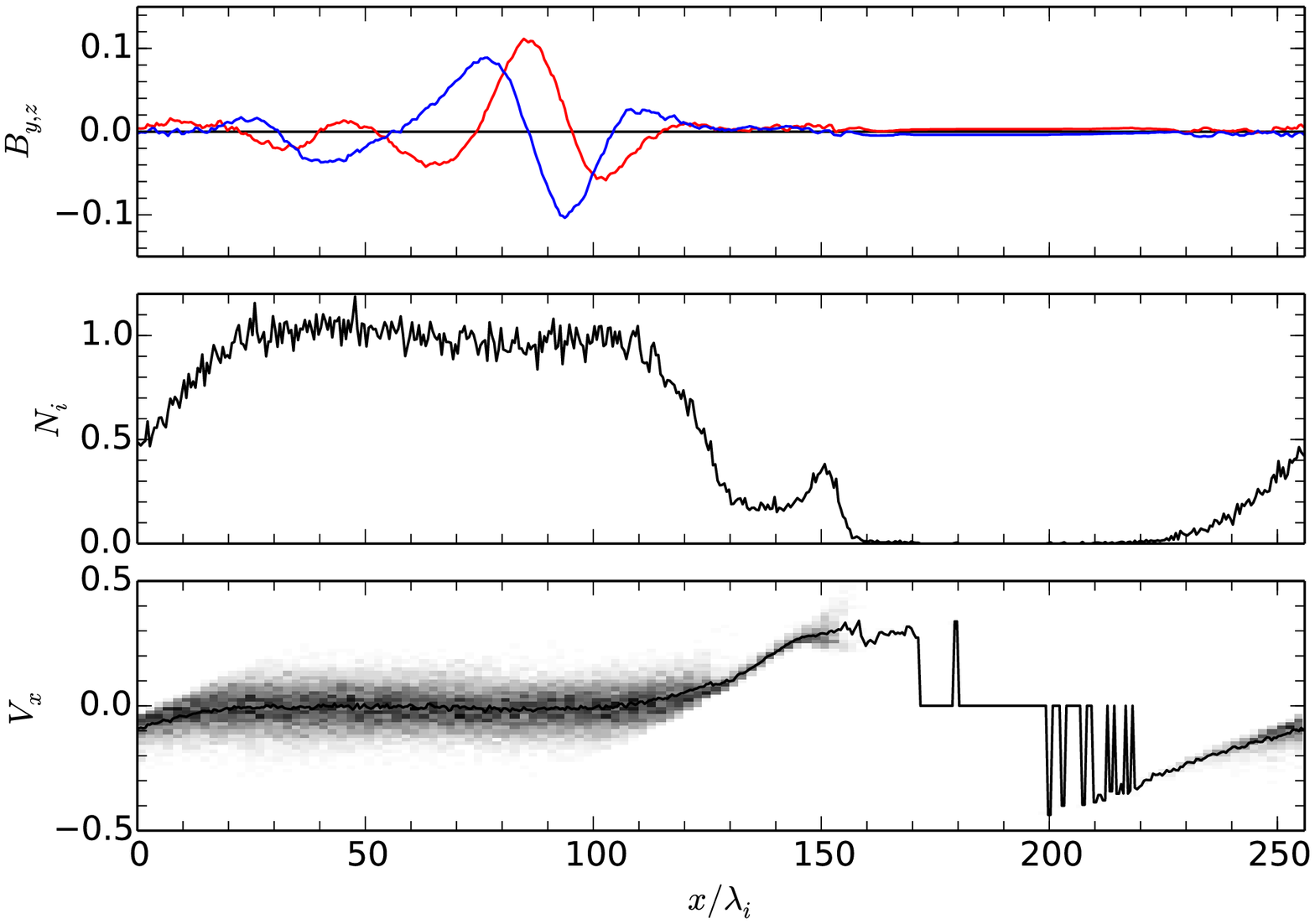}

 \caption{Snapshot of plasma expansion to vacuum with localized wave packet at
 $\Omega_{ci} t = 140.0$. The transverse magnetic field $B_y$ (red), $B_z$
 (blue) is shown in the top panel, whereas density and ion phase space diagram
 and bulk velocity $V_x$ are shown in the middle and bottom panels,
 respectively.}

 \label{fig:expansion-wave-snapshot}
\end{figure}

We have tested the effectiveness of this method by initially setting up
non-zero current density by hand in a vacuum region, which shows monotonic and
rapid decay of the initial current density as expected (not shown). Results of
another nontrivial example are shown in Figures
\ref{fig:expansion-wave-evolution} and \ref{fig:expansion-wave-snapshot}. The
setup of this simulation is the same as the previous one except that a
localized Alfv\'enic wave packet propagating in the positive $x$ direction
initially exists in the plasma region. The thickness of Gaussian envelope of
the packet is $10$ with a wave number of $k_0 = 0.196$. The maximum amplitude
of the wave packet is 20\% of the background field and is therefore large
enough for nonlinear effects being visible. Time evolution of $B_y$ and
density is shown in Fig.~\ref{fig:expansion-wave-evolution}. The wave packet
initially propagates to the right and is then reflected off the
interface. During its interaction with the interface, the density hump has
formed due to a ponderomotive force exerted by the wave packet as seen in the
left panel of Fig.~\ref{fig:expansion-wave-evolution}. This is clearly a
nonlinear effect and cannot be seen in smaller amplitude cases, which
demonstrates that the present method is robust and stable even in the presence
of nonlinear perturbations. A snapshot of the transverse magnetic field $B_y$,
$B_z$ and density and phase space diagram for ions at $\Omega_{ci} t = 140.0$
are shown in Fig.~\ref{fig:expansion-wave-snapshot}. The wave packet is
already reflected at this time and is propagating to the left. In the vacuum
region, there exists a small but finite transverse magnetic fields. The
magnetic field in vacuum is, however, constant or current-free due to the
imposed damping. Although the relaxation time to the assumed state is finite,
we think it is practically fast enough, and the relative simplicity is an
advantage of our approach.

Note that the simulation can run without introducing resistivity in this
particular case, although it gives short wavelength noise both in the plasma
and vacuum regions generated during the interaction between the wave packet
and the interface. Thus, the ``vacuum resistivity'' is not strictly necessary,
but is probably better to be included for numerical stability.

\section{Summary and Conclusions}
\label{summary}

In the present paper, we have introduced a new equation to determine the
electric field for the hybrid simulations for collisionless plasmas. The
equation takes into account finite electron inertia effect and reduces to
Laplace's equation in the limit of low density. This is in clear contrast to
the methods proposed so far that consider a correction only to the magnetic
field. This difference resolves the fundamental difficulty inherent in the
conventional hybrid simulation model, i.e, the impossibility of handling
vacuum regions due to the existence of the division-by-density operation in
the simulation procedure.

The present method improves numerical stability even for a region of finite
density. This is because the inclusion of finite electron inertia imposes a
limit of the maximum phase velocity of the system. In addition, it is
sometimes needed for numerical stability to introduce a variable mass ratio
technique, which ensures the stability by reducing the ion-to-electron mass
ratio in regions where the maximum wave phase velocity violates the CFL
stability condition.  One may think that the electron inertia effect is merely
a numerical stabilization factor rather than physics. This will hold in
most of situations where the hybrid simulation applies and the electron
inertia effect is not expected to be important, or not of primary
interest. The proposed method will thus be useful for application of the
hybrid code to problems where scale length comparable or longer than ion
inertia length is essential, while appearance of low density regions as a
result of self-consistent evolution of the system is unavoidable. It is worth
mentioning that the method does not deteriorate the advantages of the standard
hybrid code. In addition, we think that most of discussion presented in this
paper will also apply to the Hall-MHD code as well.

Finally, we note that although the current method introduces finite electron
inertia, its application to problems where electron scale physics plays a role
must be done with care, because the assumption of a fluid electron is not
always appropriate for phenomena with scale length on the order of the
electron inertia length encountered in space and astrophysical plasmas, unless
electrons are sufficiently cold.

\section*{Acknowledgement}
T.~A. thanks M. Hoshino, and T. Terasawa for useful discussion. This work was
supported by JSPS Grant-in-Aid for Young Scientists (B) 25800101.

\appendix

\section{Alternative Derivation of Equation (\ref{eq:efld_correct})}
\label{appendix-efld}

By taking temporal derivative of Ampere's law and using Faraday's law, we
obtain the following equation:
\begin{eqnarray}
 - c^2 \mathbf{\nabla} \times \mathbf{\nabla} \times \mathbf{E} =
  4 \pi \frac{\partial}{\partial t} \mathbf{J}.
  \label{eq:faraday-ampere}
\end{eqnarray}
The total current density is defined as
\begin{eqnarray}
 \mathbf{J} = \sum_s q_s \int \mathbf{v} f_s(\mathbf{v}) d \mathbf{v},
\end{eqnarray}
where $q_s$, $f_s(\mathbf{v})$ is charge and distribution function of particle
species $s$, and the sum is taken over all particle species. Now we assume
that the charged particles are collisionless and traveling under the action of
electromagnetic fields. Then, the distribution function $f_s$ obeys the Vlasov
equation
\begin{eqnarray}
 \frac{\partial}{\partial t} f_s +
  \mathbf{v} \cdot \frac{\partial}{\partial \mathbf{x}} f_s +
  \mathbf{a}_s \cdot \frac{\partial}{\partial \mathbf{v}} f_s = 0,
\end{eqnarray}
where the acceleration is given by the Lorentz force
\begin{eqnarray}
 \mathbf{a}_s = \frac{q_s}{m_s}
  \left( \mathbf{E} + \frac{\mathbf{v}}{c} \times \mathbf{B} \right).
\end{eqnarray}

One may now rewrite the right-hand side of Eq.~(\ref{eq:faraday-ampere}) as
\begin{eqnarray}
 4\pi \frac{\partial}{\partial t} \mathbf{J} &=&
  \sum_{s} 4 \pi q_s \int \mathbf{v} \frac{\partial}{\partial t} f_s d \mathbf{v}
  \nonumber \\
  &=&
   - \sum_{s} 4 \pi q_s \int \mathbf{v}
   \left[
    \mathbf{v} \cdot \frac{\partial}{\partial \mathbf{x}} f_s +
    \mathbf{a}_s \cdot \frac{\partial}{\partial \mathbf{v}} f_s
   \right] d \mathbf{v}
   \nonumber \\
 &=& - \sum_{s}
  \left[
   \mathbf{\nabla} \cdot
   \left( 4 \pi q_s \int \mathbf{v} \mathbf{v} f_s d \mathbf{v} \right) -
   4 \pi q_s \int \mathbf{a}_s f_s d \mathbf{v}
  \right]
  \nonumber \\
 &=& \sum_{s}
  \left[
   \Lambda_s \mathbf{E} + \frac{\mathbf{\Gamma}_s}{c} \times \mathbf{B} -
   \mathbf{\nabla} \cdot \mathbf{\Pi}_s
  \right],
\end{eqnarray}
where $\Lambda_s$, $\mathbf{\Gamma}_s$, $\mathbf{\Pi}_s$ are defined by
moments of the distribution function
\begin{eqnarray}
 \Lambda_s &\equiv& \frac{4 \pi q_s^2}{m_s} \int f_s d \mathbf{v} \\
 \mathbf{\Gamma}_s &\equiv& \frac{4 \pi q_s^2}{m_s}
  \int \mathbf{v} f_s d \mathbf{v} \\
 \mathbf{\Pi}_s &\equiv& 4 \pi q_s
  \int \mathbf{v} \mathbf{v} f_s d \mathbf{v}.
\end{eqnarray}
Note that, in the above derivation, we made use of the fact that
$\mathbf{\nabla}_{\mathbf{v}} \cdot \mathbf{a}_s = 0$ for the Lorentz force,
and assumed that the distribution function is a rapidly decaying function of
velocity: i.e., $f_s \rightarrow 0$ for $\mathbf{v} \rightarrow \pm \infty$,
but otherwise everything is exact.

It is easy to see that $\Lambda_s$ and $\mathbf{\Gamma}_s$ is inversely
proportional to the mass $\propto 1/m_s$, and the dominant contribution comes
from electrons. The tensor $\mathbf{\Pi}_s$ may be rewritten as
\begin{eqnarray}
 \mathbf{\Pi}_s = 4 \pi q_s
  \left(  n_s \mathbf{V}_s \mathbf{V}_s + \frac{1}{m_s} \mathbf{P}_s \right),
\end{eqnarray}
where $\mathbf{V}_s$ and $\mathbf{P}_s$ are the bulk velocity and pressure
tensor, respectively. Therefore, the contribution of the pressure gradient is
greater from electrons than ions, unless the ion-to-electron temperature ratio
is unusually high. The contribution of the term proportional to $\mathbf{V}_s
\mathbf{V}_s$ is usually small, but for electrons, it may not always be
ignored because the electron bulk velocity can become as high as the electron
\Alfven speed.

Consequently, for typical problems to which the hybrid simulation is applied,
it is sufficient to take into account the electron contribution $\Lambda_e$,
$\mathbf{\Gamma}_e$, $\mathbf{\Pi}_e$ to Eq.~(\ref{eq:faraday-ampere}). Then
we arrive at
\begin{eqnarray}
 \frac{1}{4 \pi}
  \left(
   \omega_{pe}^2 +
   c^2 \mathbf{\nabla} \times \mathbf{\nabla} \times
  \right) \mathbf{E} =
  \frac{e}{m_e}
  \left(
   \frac{\mathbf{J}_e}{c} \times \mathbf{B} - \mathbf{\nabla} \cdot \mathbf{P}_e
  \right) +
  \mathbf{\nabla} \cdot \left( \mathbf{V}_e \mathbf{J}_e \right).
\end{eqnarray}
If we further assume the charge neutrality condition $n_i \approx n_e$, we
have $\mathbf{\nabla} \cdot \mathbf{E} \approx 0$, and $\mathbf{\nabla} \cdot
\mathbf{V}_e \approx 0$, and Eq.~(\ref{eq:efld_correct}) results. Note that
the latter condition comes from the fact that temporal and spatial derivatives
of the electron density are small on the electron scale, $\partial n_e /
\partial t \approx \partial n_e / \partial x \approx 0$.

\section{Validity Range of Variable Mass Ratio}
\label{appendix-me}

In the present paper, we have introduced a technique to reduce the maximum
phase velocity of the system by locally and temporarily reducing the
ion-to-electron mass ratio. We think this is not likely to affect the
simulation results as far as the ion dynamics is concerned as explained below.

In the linear approximation for a homogeneous plasma, it is possible to
estimate the range in which finite electron inertia does not change the ion
dynamics. For this, we consider whistler waves propagating parallel to the
ambient magnetic field as it gives the maximum frequency in the system which
is the most sensitive mode to the electron dynamics. The dispersion relation
for parallel propagating right-handed circularly polarized electromagnetic
waves in a cold electron-ion plasma is given by
\begin{eqnarray}
 1 + \left( \frac{\omega_{pi}}{k c} \right)^2
  \frac{\omega}{\omega + \Omega_{ci}} +
  \left( \frac{\omega_{pe}}{k c} \right)^2
  \frac{\omega}{\omega - \Omega_{ce}}= 0
\end{eqnarray}
where $\omega/k c \ll 1$ is assumed and the cyclotron frequency $\Omega_{cs}
\, (s=i, e)$ is defined as an absolute value. For $m_e/m_i \ll 1$, it may be
approximated as
\begin{eqnarray}
 \frac{\omega^2}{(\omega + \Omega_{ci}) \Omega_{ci}} \simeq
  \left( 1 - \frac{m_e}{m_i} \frac{\omega}{\Omega_{ci}} \right)
  \left( \frac{k c}{\omega_{pi}} \right)^2
\end{eqnarray}
In the limit of $m_e \rightarrow 0$, this reduces to the usual whistler mode
branch. Assuming that the finite electron inertia becomes important at high
frequency regime ($\omega \propto k^2$), a critical wavenumber $k_c$ beyond
which the electron inertia introduces $O(\Omega_{ci})$ correction to the
frequency may be estimated as $k_c c/\omega_{pi} \simeq (m_i/m_e)^{1/4}$. The
critical wavenumber depends only weakly on mass ratio, implying that the
actual value of mass ratio is not important as far as $m_e/m_i \ll 1$. In
other words, if one wants to model a specific phenomenon correctly up to $k <
k_{max}$, the range of mass ratio required for this may be determined as
$m_i/m_e \gtrsim (k_{max} c / \omega_{pi})^4$. The dynamics of ions will not
be affected by the artificial modification of the wave dispersion as far as
this condition is satisfied.

Similar analysis for the case with nonlinear and/or inhomogeneous effects is
not easy in general, but the above condition can be used as a rough
measure. Another thing that we do not taken into account in the above analysis
is ion kinetic effect such as cyclotron resonance. Concerning the whistler
mode, this is justified unless the ion thermal velocity is much larger than
the \Alfven velocity, since otherwise the ions cannot resonantly interact with
such high frequency waves. For a very high beta plasma such that the ion
plasma beta is comparable or larger than the mass ratio ($\beta \gtrsim
m_i/m_e$), the ion cyclotron damping may not be negligible even for the
whistler branch. Nevertheless, since the general trend is to reduce the real
frequency so that the electron inertia becomes less important, one can expect
that the above conclusion will roughly hold.










\end{document}